\documentclass[letter]{aa}  

\usepackage{graphicx}
\usepackage{txfonts}
\usepackage{xcolor}
\usepackage{soul}
\usepackage{chemformula}
\usepackage{dblfloatfix}
\usepackage{multirow}
\begin{document}

   \title{Water transport through mesoporous amorphous-carbon dust}

   \author{R. Basalg{\`e}te
          \and
          G. Rouill{\'e}
          \and
          C. J{\"a}ger
          }

   \institute{Laboratory Astrophysics Group of the Max Planck Institute for Astronomy at the Friedrich Schiller University Jena, Institute of Solid State Physics, Jena, Germany\\
   \email{romain.basalgete@uni-jena.de}
    }

   \date{Received xx; accepted xx}

 
  \abstract
   {The diffusion of water molecules through mesoporous dust of amorphous carbon (a-C) is a key process in the evolution of prestellar, protostellar, and protoplanetary dust, as well as in that of comets. It also plays a role in the formation of planets. Given the absence of data on this process, we experimentally studied the isothermal diffusion of water molecules desorbing from water ice buried at the bottom of a mesoporous layer of aggregated a-C nanoparticles, a material analogous to protostellar and cometary dust. We used infrared spectroscopy to monitor diffusion in low temperature (160 to 170~K) and pressure (6~$\times$ 10$^{-5}$ to 8~$\times$ 10$^{-4}$~Pa) conditions. Fick's first law of diffusion allowed us to derive diffusivity values on the order of 10$^{-2}$~cm$^2$~s$^{-1}$, which we linked to Knudsen diffusion. Water vapor molecular fluxes ranged from 5~$\times$ 10$^{12}$ to 3~$\times$ 10$^{14}$~cm$^{-2}$~s$^{-1}$ for thicknesses of the ice-free porous layer ranging from 60 to 1900~nm. Assimilating the layers of nanoparticles to assemblies of spheres, we attributed to this cosmic dust analog of porosity 0.80--0.90 a geometry correction factor, similar to the tortuosity factor of tubular pore systems, between 0.94 and 2.85. Applying the method to ices and refractory particles of other compositions will provides us with other useful data.}


   \maketitle
%

\section{Introduction}

To date, the soil of comets that have already known activity appears to consist of porous dust that contains water ice in its depths \cite[e.g.,][and references therein]{Lisse22}. During the active phase of comets, water somehow diffuses through the porous cometary soil and escapes to contribute to the gas coma \citep[e.g.,][]{lecacheux_observations_2003, kelley_spectroscopic_2023}. The mechanism of this diffusion or transport is the object of experiments \citep[e.g.,][]{gundlach_outgassing_2011, schweighart_viscous_2021} and modeling \citep[e.g.,][and references therein]{skorov_activity_2011, Laddha23, Guettler23}.

We are also interested in the outgassing of porous, refractory materials that contain ices. Aggregated grains of refractory materials that have accumulated ices are components of the dust observed in prestellar cores, as well as in protostellar and protoplanetary disks \citep[e.g.,][]{Boogert_2015}. The refractory materials are mainly silicates and carbonaceous matter, while the ices consist essentially of water (H$_2$O) \citep[e.g.,][]{Demyk_2011}. The sublimation of the volatile icy content of the dust and the associated outgassing cause, for example, ice lines and spatial variations of the gas-to-dust mass ratio \citep[e.g.,][]{Stammler17, Schoonenberg_2017, Spadaccia_2021}. Thus, they affect the trajectory of dust particles and, eventually, planet formation.

The outgassing of dusts lacks experimental characterization. Consequently, the modeling of dust in disks ignores its porosity and its effect on outgassing \citep[e.g.,][]{Stammler17}. To solve this issue, we use our experience in producing layers of aggregated amorphous, refractory nanoparticles \citep{Jaeger_2008, Sabri_2014} that are analogs of the interstellar dusts from which prestellar, protostellar, and protoplanetary dusts have formed. We present a new experimental method that allows us to quantify the transport of volatiles in these materials and to derive physical parameters such as diffusion coefficients. We demonstrate the method with the case of water molecules desorbing from buried ice and diffusing through mesoporous layers of amorphous-carbon (a-C) nanoparticles.

\section{Experimental}\label{sec:exper}

\subsection{Preparation of a-C grains and water ice layers}\label{sec:exper1}

\begin{figure}
\centering
\resizebox{8.8cm}{!}{\includegraphics{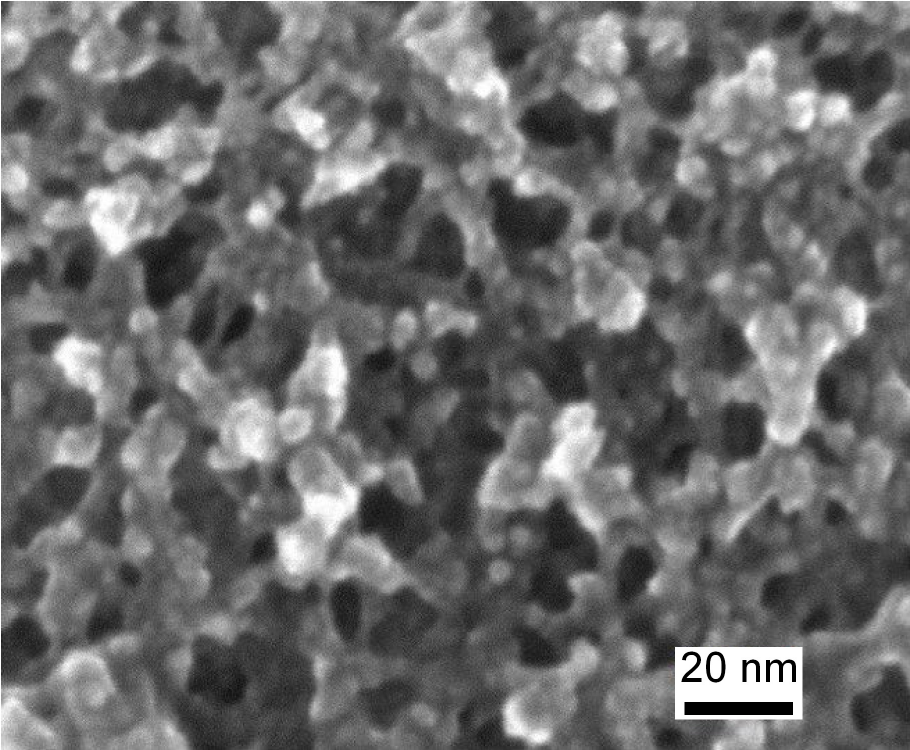}}
\caption{FESEM image of a porous layer of aggregated a-C grains produced for this study.}
\label{fig:FESEM}
\end{figure}

We synthesized a-C nanoparticles using laser ablation of a graphite target under a quenching atmosphere of He gas \citep{Jaeger_2008}. Appendix~\ref{sec:synth} offers detailed information on the procedure. The synthesis apparatus allowed us to generate a beam of these analogs of cosmic carbon grains, to deposit layers of them on KBr substrates, and to perform IR transmission spectroscopy without breaking the vacuum. Figure~\ref{fig:FESEM} shows a field emission scanning electron microscopy (FESEM) image of a deposit produced for this study. Individual a-C grains, $\sim$1 to $\sim$8~nm in diameter, went on to form aggregates and their accumulation produced a mesoporous layer. Its structure is similar to that of a random ballistic deposit of particles, which is not surprising given the method used for its preparation (Appendix~\ref{sec:synth}). The diameter of the pores follow a distribution from a few nanometers up to 100~nm, with a typical average on the order of 10--20~nm (mesopores). The porosity of the material, denoted by $\epsilon$, ranged from 0.80 to 0.90.

We used a quartz crystal microbalance (QCM) to monitor the deposition of grains and measure the thickness $L_\mathrm{C}$ of the deposited layer, ignoring porosity, detailed in Appendix~\ref{sec:Lc-IR}. The derivation of $L_\mathrm{C}$ takes into account a density of 1.55~g~cm$^{-3}$ for the a-C material \citep{Jaeger_2008}. We determined the thickness $L$ of a porous deposit according to:
\begin{equation}\label{eq:Lepsilon}
   L = \frac{L_{\mathrm{C}}}{(1-\epsilon)} .
\end{equation}

The synthesis apparatus also made it possible to add a flow of water vapor, H$_2$O(g), parallel to the beam of nanoparticles and a cryostat equipped with a heating element stabilized the substrates to the low temperatures required by the study. Thus, we could deposit grains on a cold substrate together with H$_2$O molecules so as to obtain a composite layer of a-C nanoparticles and water ice, H$_2$O(s). We evaluated the amount of deposited water in terms of molecular column density, denoted $N_{\mathrm{H}_2\mathrm{O(s)}}$, using the area of the 3~$\mu$m absorption band appearing in the IR spectra and the relevant band strength (see below).

\subsection{Isothermal desorption-diffusion experiments}\label{sec:exper2}

\begin{figure}
\centering
\resizebox{8.8cm}{!}{\includegraphics{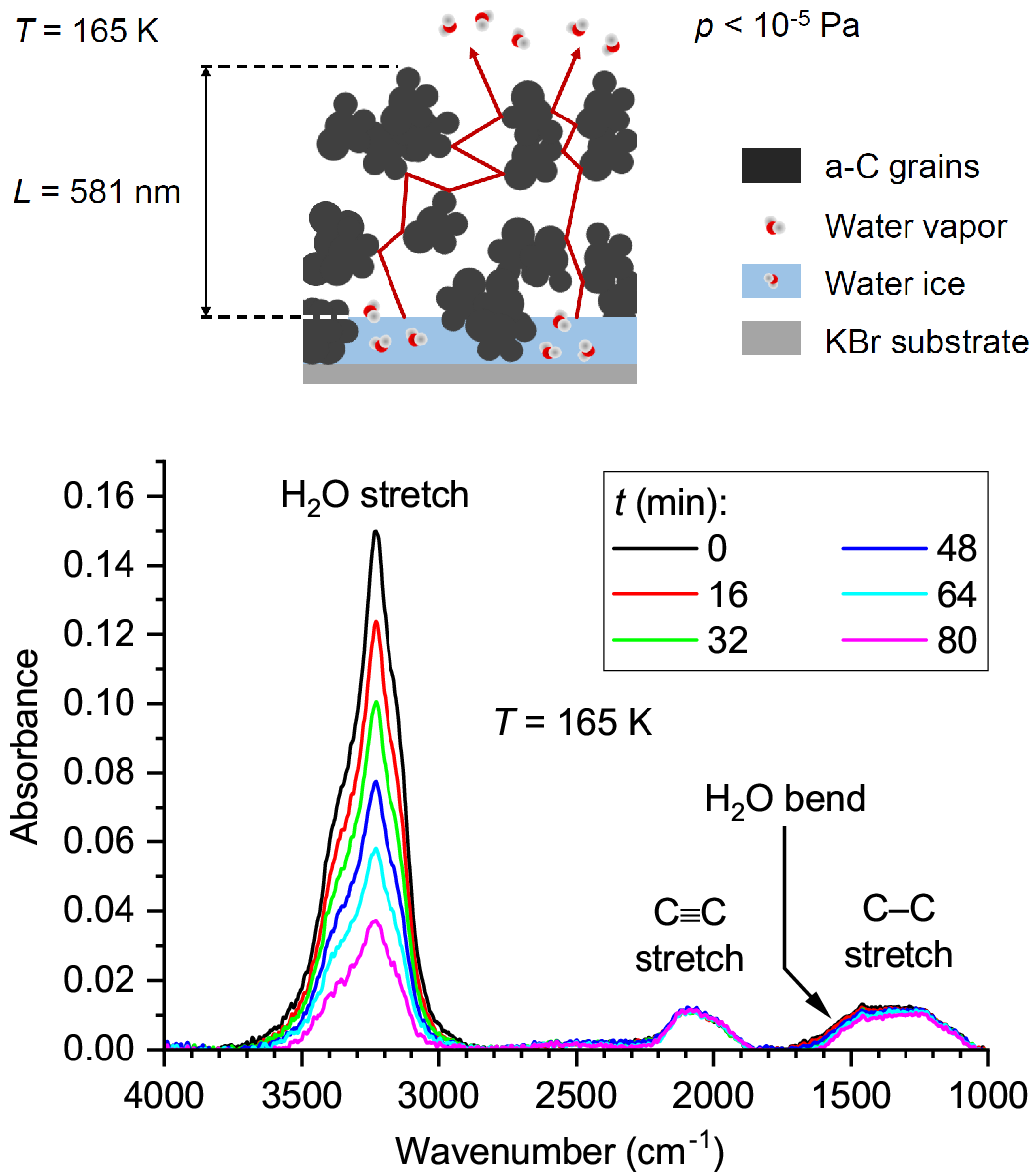}}
\caption{Isothermal time evolution of the IR spectrum of a deposit of water ice and a-C grains. The temperature $T$ of the substrate was 165~K and the thickness $L$ of the a-C grain layer was 581~nm. Top panel: Sketched cross-section of the structure of the deposit with H$_2$O molecules diffusing through the porous material after desorption from the ice. Bottom panel: Spectrum with absorption bands of crystalline H$_2$O(s) and a-C grains with indication of the vibrational modes.}
\label{fig:IR}
\end{figure}

We conducted experiments on the desorption and diffusion of H$_2$O molecules as follows. (i) We first deposited a-C grains and H$_2$O molecules at the same time on a KBr substrate kept at 150~K. Thus, a composite layer of grains and ice covered the substrate. The a-C material amounted to an  $L_\mathrm{C}$ value of 3 to 6~nm depending on the experiment. At 150~K, H$_2$O molecules formed ice in the hexagonal crystalline polymorph form \citep{Jenniskens_1998}. Figure~\ref{fig:IR} shows a spectrum to which the composite layer contributes and the profile of the 3~$\mu$m absorption band confirms the phase state of water. We measured the area of the band and, taking into account a strength (or absorption length) of 2.7~$\times$ 10$^{-16}$~cm \citep{Mastrapa_2009}, we derived the column density of the deposited H$_2$O molecules. Depending on the experiment, it was equivalent to a number of monolayers in the 70--200 range, hence, to a thickness of 22 to 64~nm, considering a molecular monolayer density of 10$^{15}$~cm$^{-2}$ and a specific mass of 0.931~g cm$^{-3}$ \citep{Kuhs87,Mastrapa_2009}. Assuming the composite layer did not comprise empty volumes, its compounded thickness was thus in the 25--70~nm range. (ii) On top of the composite layer, still at 150~K, we deposited a-C grains until we obtained a layer of the desired thickness $L$, typically in the 60--1900~nm range, about 2 to 30 times thicker than the composite layer. Thus, the procedure produced a composite layer of aggregated a-C grains and water ice covered with a thicker, ice-free, mesoporous layer of similar a-C grains (see sketch in Fig.~\ref{fig:IR}). (iii) We then heated the substrate at a rate of 10~K~min$^{-1}$ and stabilized the temperature at either 160, 165, or 170~K in separate experiments. These temperatures being higher than the desorption temperature of H$_2$O molecules from water ice \citep{fraser_thermal_2001, bolina_2005}, they enabled the desorption and diffusion of the molecules. After the substrate temperature and the chamber pressure stabilized, we measured the area of the H$_2$O(s) 3~$\mu$m band as a function of time. The area decreased, as Fig.~\ref{fig:IR} shows for a substrate at 165~K and an a-C grain layer of thickness $L$~= 581~nm.

\section{Results}\label{sec:results}

\subsection{Isothermal experiments and diffusion coefficients}

\begin{figure*}
\centering
\resizebox{18.4cm}{!}{\includegraphics{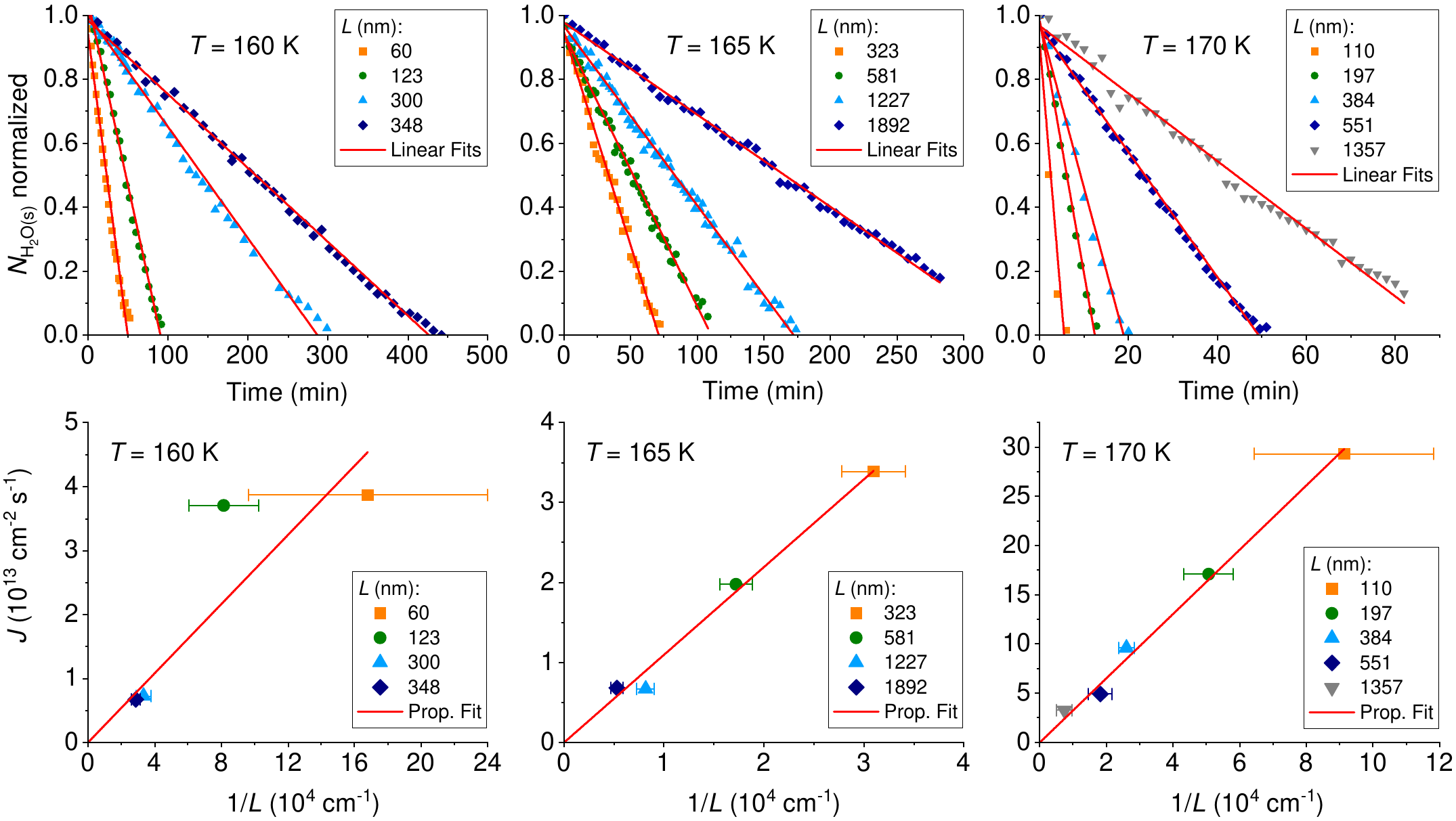}}
\caption{\textbf{Experimental results.} Time evolution of $N_{\mathrm{H}_2\mathrm{O(s)}}$, normalized to its initial value, during isothermal experiments at 160~K (left), 165~K (middle), and 170~K (right) shown in the top panel, for several values of $L$, the thickness of the top layer of a-C grains. Bottom panels show the Molecular flux $J_{\mathrm{H}_2\mathrm{O}}$ derived from proportional fits of the linear decays of $N_{\mathrm{H}_2\mathrm{O(s)}}$ reported in the top panels, plotted as a function of $1/L$. Error bars reflect the uncertainty introduced by $\epsilon$ and $L_\mathrm{C}$ (see Appendix~\ref{sec:Lc-IR}).}
\label{fig:results}
\end{figure*}

We carried out the isothermal experiments at three temperatures, specifically 160, 165, and 170~K, for several thicknesses, $L,$ of the top, ice-free layer of a-C grains. Figure~\ref{fig:results} summarizes the results. The top panels display the time evolution of $N_{\mathrm{H}_2\mathrm{O(s)}}$ during the experiments. For clarity, we normalize each series of column densities to the value measured at time $t$~= 0. The curves show that the column density decreased at a constant rate during every experiment and that the rate depended on the temperature and on $L$. We do not observe any variation of the shape of the 3~$\mu$m band during the experiments, an indication that condensed water was present at all times only as crystalline ice. This allowed us to take into account a constant band strength (Sect.~\ref{sec:exper2}) to derive $N_{\mathrm{H}_2\mathrm{O(s)}}$ at all times of every experiment. In most of them, we measured spectra until the ice completely disappeared.

The decrease of $N_{\mathrm{H}_2\mathrm{O(s)}}$ probes the amount of H$_2$O molecules desorbing from the ice of the composite layer, that is, from ice located at the bottom of a porous layer of a-C grains. The desorbed molecules left the porous layer because of the surrounding constant vacuum and, given that the lateral dimensions of the layer were three orders of magnitude greater than $L$, we can define the molecular flux $J_{\mathrm{H}_2\mathrm{O}}$ through the layer with:
\begin{equation}\label{eq:dNdt}
   J_{\mathrm{H}_2\mathrm{O}}=-\frac{\mathrm{d}N_{\mathrm{H}_2\mathrm{O(s)}}(t)}{\mathrm{d}t} ,
\end{equation}
where $J_{\mathrm{H}_2\mathrm{O}}$ is not time dependent since the curves in the top panels of Fig.~\ref{fig:results} show that $N_{\mathrm{H}_2\mathrm{O(s)}}(t)$ varied linearly with time in every experiment.

The bottom panels of Fig.~\ref{fig:results} show the values of $J_{\mathrm{H}_2\mathrm{O}}$ as a function of $1/L$ and reveal their proportionality. According to Fick's first law of diffusion \citep[for a review on diffusion processes in mesoporous materials, see][]{bukowski_connecting_2021}, the relation between $J_{\mathrm{H}_2\mathrm{O}}$ and the molecular concentration $C_{\mathrm{H}_2\mathrm{O}}$ through the porous layer is
\begin{equation}\label{dcdz}
   J_{\mathrm{H}_2\mathrm{O}}(T)=-D_{\mathrm{H}_2\mathrm{O}}(T)\frac{\mathrm{d}C_{\mathrm{H}_2\mathrm{O}}(z,T)}{\mathrm{d}z} ,
\end{equation}
where $D_{\mathrm{H}_2\mathrm{O}}(T)$ is the diffusivity or diffusion coefficient of the water molecules at temperature, $T$, and $z$ is the position or height measured from the surface of the ice, that is, the sublimation front, the location of which we considered constant owing to the lack of information on its geometry as a function of time. The assumption was otherwise a reasonable approximation in the experiments with a composite layer that was very thin compared to the ice-free a-C layer. We postulate that the concentration $C_{\mathrm{H}_2\mathrm{O}}(0,T)$ at the bottom of the a-C grain layer corresponded to the vapor pressure of H$_2$O(g) above H$_2$O(s) at the temperature, $T,$ and take the value from \citet{Mauersberger_2003}. The concentration at the top of the layer was negligible compared to $C_{\mathrm{H}_2\mathrm{O}}(0,T)$ since the pressure in the chamber was lower than 10$^{-5}$~Pa. If we now assume that the concentration gradient of H$_2$O molecules decreased linearly from the bottom to the top of the a-C grain layer, we obtain a proportionality relation between $J_{\mathrm{H}_2\mathrm{O}}$ and $1/L$, namely:\begin{equation}
   J_{\mathrm{H}_2\mathrm{O}}(T)=D_{\mathrm{H}_2\mathrm{O}}(T)\frac{C_{\mathrm{H}_2\mathrm{O}}(0,T)}{L} ,
\label{eq:J_D}
\end{equation}
which is consistent with what we observed in the experiments (Fig.~\ref{fig:results}). Consequently, fitting proportionality functions to the data reported in the bottom panels of Fig.~\ref{fig:results} gives us $D_{\mathrm{H}_2\mathrm{O}}(T)  $, with  $C_{\mathrm{H}_2\mathrm{O}}(0,T)$ already known. Table~\ref{tab:dif_r} presents the diffusion coefficients thus obtained, the uncertainties being determined in the fitting procedure. 

\begin{table}
\caption{Diffusivity of H$_2$O molecules in a mesoporous layer of aggregated a-C grains.}
\label{tab:dif_r}
\centering
\begin{tabular}{llll}
\hline\hline
$T$ & $p(0)$             & $D_{\mathrm{H}_2\mathrm{O}}=D_{\mathrm{H}_2\mathrm{O}}^{\mathrm{K}}$ & $q$  \\
(K) & (Pa)               & (10$^{-2}$~cm$^{2}$~s$^{-1}$)                                        &         \\
\hline
160 & 6~$\times 10^{-5}$ & 1.04~$\pm$ 0.15                                                      & 1.03--2.60 \\
165 & 2~$\times 10^{-4}$ & 1.14~$\pm$ 0.10                                                      & 0.94--2.37 \\
170 & 8~$\times 10^{-4}$ & 0.97~$\pm$ 0.05                                                      & 1.13--2.85 \\
\hline
\end{tabular}
\tablefoot{$T$: H$_2$O(s) temperature; $p(0)$: H$_2$O(g) pressure at the ice surface; $D_{\mathrm{H}_2\mathrm{O}}$: measured diffusion coefficient for H$_2$O molecules; $D_{\mathrm{H}_2\mathrm{O}}^{\mathrm{K}}$: Knudsen diffusion coefficient for H$_2$O molecules; and $q$: correction factor for the geometry of the layer.}
\end{table}


\subsection{Diffusion process and characterization of the mesoporous a-C dust}\label{sec:astro}

\citet{bukowski_connecting_2021} discussed diffusion mechanisms in mesoporous materials in detail. In brief, surface diffusion, Knudsen diffusion, and molecular diffusion can occur. Molecular diffusion proceeds through collisions between molecules. Under our experimental conditions, H$_2$O(g) was the only substance that needed to be taken into account. As its pressure was very low ($p$~$<$ 10$^{-3}$~Pa), the mean free path of the molecules was much greater than the dimensions of the pores (Knudsen number~$\gg$ 1) and molecular diffusion could not occur. The question is which of the surface diffusion and Knudsen diffusion dominated. Surface diffusion is an activated process for which the diffusivity follows an Arrhenius-type law, that is, $D^{\mathrm{S}}~\propto$ $\mathrm{e}^{-E_{\mathrm{A}}/k_{\mathrm{B}}T}$, where $E_{\mathrm{A}}$ is the activation energy and $k_{\mathrm{B}}$ is the Boltzmann constant. Knudsen diffusion is an effect of the molecules colliding with the walls of the porous material and the dependence of the Knudsen diffusivity on temperature is such that $D^{\mathrm{K}}$~$\propto$ $T^{1/2}$.
Therefore, if one mechanism dominated the other in the experiments, the diffusion coefficients varied following the relevant temperature dependence. However, the uncertainty that affects the $D_{\mathrm{H}_2\mathrm{O}}(T)$ values that we derived for temperatures in the 160--170~K range does not allow us to reject one of the mechanisms. Nevertheless, we argue that (i) the apolar nature of the a-C material results in weak interactions of the H$_2$O molecules with its surface; (ii) then, if surface diffusion occurred, the bands of H$_2$O molecules adsorbed at favorable binding sites would be displayed in the spectra at the end of the isothermal experiments, after a total depletion of the ice, however, this is not the case; and (iii) the modeling of the diffusion with a Knudsen regime gave reasonable results (see below). Thus, we find that Knudsen diffusion dominated the transport of H$_2$O molecules in the experiments.

In order to characterize the structure of the present a-C grain layers, we assimilate them to assemblies of monodisperse spheres as those formed by random ballistic deposition. Following \citet{Asaeda74}, in a medium of porosity $\epsilon$ that consists of packed spheres with an average diameter $<$\,$a$\,$>$, the Knudsen diffusivity, $D^{\mathrm{K}}$, is such that:
\begin{equation}
   D^{\mathrm{K}} = \frac{1}{3} <a> \frac{\epsilon^2}{1-\epsilon} \frac{1}{q\Phi} \sqrt{\frac{8RT}{\pi M}} ,
\label{eq:dk2}
\end{equation}
where $R$ is the universal gas constant, $M$ is the molar mass of the diffusing gas, $q$ is a correction factor for the geometry of the packing, similar in role to the tortuosity factor of tubular pores, and $\Phi$ is a constant equal to 2.18 (or 13/6 in \citealt{Derjaguin46} and \citealt{Guettler23}; we note that \citealt{Mekler90} modeled the material of cometary nuclei with randomly packed spheres to study ice sublimation in these objects). In this study, we observed that $a$ ranged from $\sim$1 to $\sim$8~nm and estimated that $\epsilon$ was 0.80--0.90 (Sect.~\ref{sec:exper1}). Setting $<$\,$a$\,$>$ to 5~nm and having measured the Knudsen diffusivity of water molecules $D_{\mathrm{H}_2\mathrm{O}}^{\mathrm{K}}$ as $D_{\mathrm{H}_2\mathrm{O}}$, we obtained the missing descriptor $q$ using Eq.~(\ref{eq:dk2}). Table~\ref{tab:dif_r} presents the values obtained for $q$ and they verify the condition:
\begin{equation}
   \frac{3 (1 - \epsilon)}{\epsilon <a>} \frac{\Phi q L}{4} \gg 1
\label{eq:condition}
,\end{equation}
which \citet{Asaeda74} used to write Eq.~(\ref{eq:dk2}). We note that the values, which range from 0.94 to 2.85, encompass the value of 1.41 determined by \citet{Asaeda74} with packed beds of glass spheres and those derived by \citet{Guettler23} with similar media. Since we observed diffusion in Knudsen regime, we evaluated an effective description of the a-C layer with a medium featuring tubular pores in Appendix~\ref{sec:q-and-tau}.

\section{Discussion and conclusion}

The gas-phase condensation that we exploited to synthesize a-C nanoparticles is comparable to the formation mechanism of grains in late-type stars \citep{Jaeger_2008}. The deposited mesoporous layers of aggregated particles are analogues of interstellar a-C dust, hence, of the carbon component of prestellar, protostellar, and protoplanetary dusts. The a-C material is also relevant to cometary studies as it shows similarities with dust particles in the ejecta of comets. For example, spectral observations of the ejecta of comet 9P/Tempel~1 revealed the presence of a-C matter \citep{Lisse_2006}. In terms of morphology, the Rosetta mission collected fluffy dust with porosity up to 0.85 \citep[][and references therein]{levasseur-regourd_cometary_2018}. 

Past experimental studies have focused on the spectral properties of the synthesized a-C materials \citep{Jaeger_2008, Jaeger_2009} and molecular diffusion through them lacked characterization. At 150~K in vacuum, we produced mesoporous layers of aggregated synthetic a-C nanoparticles on top of a mixture of the same particles and crystalline water ice. The average particle size was 5~nm and the porosity 0.80 to 0.90. Stabilizing the temperature of the system in the 160--170~K range, we observed the disappearance of the ice as a function of time with IR spectroscopy and concluded that the ice sublimated and the system outgassed. The analysis of the spectra revealed the kinetic of the phenomenon and we derived molecular fluxes at three temperatures for several thicknesses of the top layer of a-C grains. With indications that Knudsen diffusion occurred, we used Fick's first law of diffusion to determine diffusion coefficients for H$_2$O molecules in the mesoporous a-C layer, which are found in the order of 10$^{-2}$ cm$^2$ s$^{-1}$. At 160 to 170~K and for mesoporous a-C layers 60 to 1900~nm thick, the outgassing ranged from 5~$\times$ 10$^{12}$--3~$\times$ 10$^{14}$~cm$^{-2}$ s$^{-1}$ in terms of molecular flux. We could further characterize the mesoporous a-C layers with a geometry correction factor between 0.94 and 2.85.

The immediate implication of our experiments is that a dust grain that is a ballistic aggregate of a-C nanoparticles that has accumulated water ice would rapidly lose its entire H$_2$O(s) content when crossing the water ice line in an accretion disk. Actual interstellar carbonaceous grains would be partially hydrogenated, possibly functionalized and the diffusion of H$_2$O molecules in the pores of these active-carbon grains may show significant differences compared to the pure a-C grains studied here. Concerning silicates, their surface strongly interacts with water molecules and features silanol groups (\ch{Si\bond{sb}OH}) and adsorbed molecules that only high temperatures can remove. Consequently, we expect this interaction to play a role in the transport of H$_2$O molecules through the pores of silicate dust. Our experimental method will allow us to study these cases and also to obtain data on other relevant volatiles.

\begin{acknowledgements}
This work received the support of the Deutsche Forschungsgemeinschaft (DFG) through project No. 504825294 "From atoms to prebiotic molecules on the surface of cosmic dust grains". The authors are grateful to C. G{\"u}ttler for comments that enabled them to improve the manuscript.
\end{acknowledgements}

%
\bibliographystyle{aa} 
\bibliography{Bibliolite} 
%

\begin{appendix}

\section{Synthesis of a-C nanoparticles and layer production}\label{sec:synth}

We synthesized a-C nanoparticles by pulsed laser ablation of a rotating graphite target under a quenching atmosphere that consisted of He gas (Westfalen, 99.999\% purity) at 4~Torr pressure. The laser source (Continuum Surelite I-10) emitted light of 532~nm wavelength in pulses of 5~ns duration and 50~mJ energy at a repetition rate of 10~Hz. A lens focused the laser light onto the target giving an energy density of 25~J cm$^{-2}$ at the target.

Differential pumping through a nozzle and a skimmer (Beam Dynamics, Inc., 1.2~mm orifice) allowed us to extract the grains, form them into a beam and deposit them in a separate vacuum chamber on a 2 mm-thick KBr substrate (Korth Kristalle GmbH) attached to the cold head of a compressed-helium, closed-cycle cryocooler (Advanced Research Systems, Inc. ARS-4H and DE-204SL).
A quartz crystal microbalance (QCM) received a portion of the grains during production of a layer. It allowed us to measure the amount of deposited material and to determine the thickness $L_\mathrm{C}$ of the layer, where $L_\mathrm{C}$ did not take porosity into account.

\section{Evaluation of $L_\mathrm{C}$}\label{sec:Lc-IR}

We used a quartz crystal microbalance (QCM) to prepare the carbon grain layers for the diffusion experiments at 165~K temperature. In order to determine $L_\mathrm{C}$ for the carbon grain layers produced for the experiments at 160 and 170~K, we compared their IR spectra with those of the layers characterized with the QCM. In each experiment, the initial IR spectrum of the first layer, composite of carbon grains and water ice, served as the reference or baseline for the spectroscopy of the second layer. Practically, we measured the areas of the absorption bands caused by the amorphous carbon material, between 900 and 1800~cm$^{-1}$ and between 1800 and 2300~cm$^{-1}$ for the \ch{C$\bond{sb}$C} and \ch{C$\bond{tp}$C} stretching vibrations, respectively. Then, taking the data of the diffusion experiments conducted at 165~K, we fitted for both types of bands a coefficient that related the areas of these bands to the corresponding $L_\mathrm{C}$ values measured with the QCM. Finally, we applied the coefficients thus derived to the areas of the bands that arose in the IR spectra of the carbon grain layers deposited for the experiments at 160 and 170~K. The difference between the results that the two types of bands gave indicates an uncertainty between 9\% and 43\% on $L_\mathrm{C}$.

\section{Effective modeling with tubular pores}\label{sec:q-and-tau}

\begin{figure}
\centering
\resizebox{8.8cm}{!}{\includegraphics{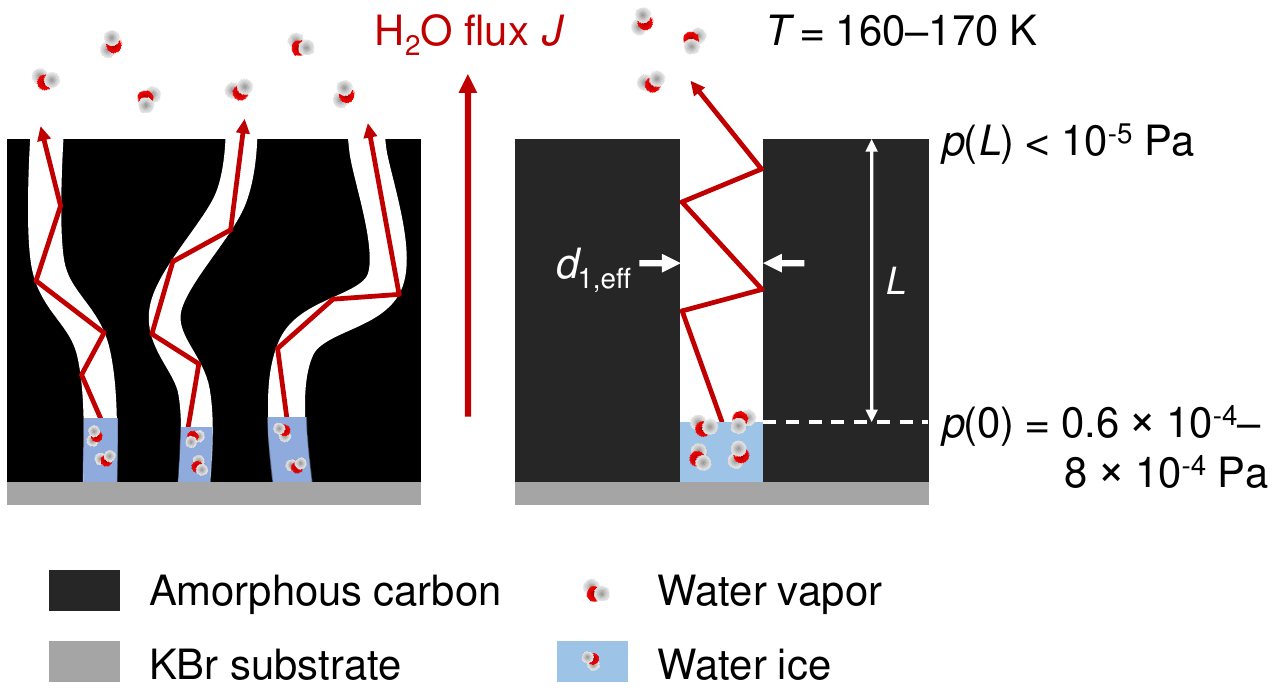}}
\caption{Modeling Knudsen diffusion of H$_2$O molecules in a mesoporous a-C layer: layer with tortuous pores characterized with the porosity, $\epsilon$, the average diameter, $<$$d$$>$, and the tortuosity factor, $\tau$ (left); layer with a single straight pore of effective diameter, $d_{\mathrm{1,eff}}$, and length, $L$ (right).}
\label{fig:model}
\end{figure}

The flows measured in this study being consistent with a Knudsen regime, we can model the a-C grain layers with an effective solid medium of porosity $\epsilon$, the pores being tubes with an average diameter of $<$$d$$>$ (left side of Fig.~\ref{fig:model}). We remark that \citet{Mekler90} also compared the two approaches in a study of ice sublimation in a cometary nucleus. Knudsen diffusion predicts that the diffusivity $D^{\mathrm{K}}$ of a gas in this medium is given by:
\begin{equation}
   D^{\mathrm{K}} = \frac{1}{3} \frac{\epsilon}{\tau} <d> \sqrt{\frac{8RT}{\pi M}} ,
\label{eq:dk3}
\end{equation}
where $\tau$ is a tortuosity factor that contains information on the geometry of the pores \citep{bukowski_connecting_2021}. Specifically, $\tau$ compares the average path length of the pores, $<$\,$L^{\prime}$\,$>$, with the thickness of the medium, $L$, following:
\begin{equation}
   \tau = \left( \frac{<L^\prime>}{L} \right)^2 .
\label{eq:tau}
\end{equation}
Given that $<$$d$$>$ and $\epsilon$ were 10--20~nm and 0.80--0.90, respectively (Sect.~\ref{sec:exper1}), entering the measured $D^{\mathrm{K}}$ values (as $D_{\mathrm{H}_2\mathrm{O}}$) into Eq.~(\ref{eq:dk3}) gave $\tau$ at 1.0--2.8. The lower value of $\tau$ cannot be smaller than 1 according to Eq.~(\ref{eq:tau}). We remark that the values obtained for $q$ with the sphere assembly model and $\tau$ are similar, suggesting that the model of tubular pores is actually effective. The comparison of Eq.~(\ref{eq:dk2}) with Eq.~(\ref{eq:dk3}) shows that the correction factor $q$ plays for assemblies of monodisperse spherical particles the role that the tortuosity factor $\tau$ plays for media with tubular pores \citep{Guettler23}.

Now considering a simpler model to describe the a-C grain layers, that is, a single straight pore of diameter $d_{\mathrm{1,eff}}$ with $d_{\mathrm{1,eff}}$~$<<$ $L$ (right side of Fig.~\ref{fig:model}), we can write
\begin{equation}
   D^{\mathrm{K}} = \frac{1}{3} d_{\mathrm{1,eff}} \sqrt{\frac{8RT}{\pi M}} .
\label{eq:dk1}
\end{equation}
With measurements of $D_{\mathrm{H}_2\mathrm{O}}$ at hand, we derived the effective pore diameter, $d_{\mathrm{1,eff}}$, using Eq.~(\ref{eq:dk1}). We obtained values between 6.0 and 8.5~nm, close to the average diameter of the actual pores $<$$d$$>$ of 10 to 20~nm. This result strengthens our conclusion that Knudsen diffusion dominated the transport of the H$_2$O molecules in our experiments.

\end{appendix}

\end{document}